\documentclass[letterpaper,11pt,reqno]{article}
\usepackage{hyperref}
\usepackage{amsfonts}
\usepackage{amsmath}
\usepackage{multicol}
\usepackage{multirow}
\usepackage{geometry}
\usepackage{graphicx}
\usepackage{subfigure} 
\usepackage[all]{xy}
\usepackage{indentfirst}
\usepackage{appendix}
\usepackage[affil-it]{authblk}
\usepackage{etoolbox}
\begin{document}
% generates the title

\title{{The Number of Rational Points of Two Parameter Calabi-Yau manifolds as Toric Hypersurfaces}}
\date{}
\author[]{Yuan-Chun Jing, Xuan Li and Fu-Zhong Yang  \thanks{corresponding author \hspace{1cm} E-mail:\href{fzyang@ucas.ac.cn}{fzyang@ucas.ac.cn}}}
\affil[]{School of Physical Sciences,University of Chinese Academy of Sciences,\\No.19(A) Yuquan Road, Shijingshan District, Beijing, P.R.China 100049}

\maketitle

\begin{abstract}
The number of rational points in toric data are given for two-parameter Calabi-Yau $n$-folds as toric hypersurfaces over finite fields $\mathbb F_p$ . 
We find that the fundamental period is equal to the number of rational points of the Calabi-Yau $n$-folds in zeroth order $p$-adic expansion. 
By analyzing the solution set of the GKZ-system given by the enhanced polyhedron, we deduce that under type II/F-theory duality the 3D and 4D Calabi-Yau manifolds have the same number of rational points in zeroth order.
Taking the quintic and its duality as an example, the number of rational points in some specific complex moduli are given by numerical calculation to support our results.
\end{abstract}

% insert the table of contents
\tableofcontents

\section{Introduction}
The period integrals of holomorphic 3-form on Calabi-Yau manifolds play an important role in topological string theory.
The mirror symmetry between two Calabi-Yau manifolds $(M_3,W_3)$ makes it possible to use the pure geometric data of $W_3$ to calculate the Yukawa coupling of the low-energy effective theory after the superstring compactification on $M_3$ \cite{1}.
The Yukawa coupling is of great interests in physics because it contains non-perturbative quantum corrections from holomorphic instantons, mathematically corresponding to Gromov-Witten invariants.

A classic result in arithmetic geometry is that the number of rational points of elliptic curves over finite field $\mathbb F_p$ can be given by periods as hypergeometric functions, as shown in \cite{2}.
Recent work \cite{3} has extended this result to higher genus curves over $\mathbb F_{p^n}$.
As a generalization of this result from elliptic curves to Calabi-Yau manifolds, \cite{4} gives the number of rational points for quintic in terms of periods in  $p$-adic expansion.
These results connect complex geometry with arithmetical geometry, and give a number theoretic meaning to the geometric quantities that superstring theory cares about.

Then, $\zeta$ functions, as the generating functions for the number of rational points, of Calabi-Yau manifolds over finite fields are studied \cite{5}, \cite{6}, \cite{7}. 
In particular, some propositions and conjecture about arithmetical mirror symmetry are given. 
Recently, the local $\zeta$ functions are investigated using the Picard-Fuchs equations satisfied for several single-parameter manifolds, and gives rather comprehensive numerical results \cite{8}.

In superstring theory , Calabi-Yau manifolds with more than one moduli are interests for physics and mathematics , such as in type II/F-theory duality \cite{9}, the dimension of moduli space of the Calabi-Yau 4-fold $W_4$ is equal to the sum of the dimension of moduli space of the Calabi-Yau 3-fold $W_3$ and its divisor $D$.
We wish to generalize the relation between periods and the number of rational points to multi-parameter Calabi-Yau manifolds.
In this article, the number of rational points $\nu_0(\phi,\psi)$ in toric data is given for two-parameter Calabi-Yau $n$-folds as toric hypersurfaces. 
We find that $\nu_0(\phi,\psi)$ is equal to the fundamental period $\varpi_0(\phi,\psi)$ of Calabi-Yau $n$-folds over finite fields in zeroth order $p$-adic expansion. 

When considering type II/F-theory duality, our results can give some insights on the number of rational points of the dual manifolds $W_4$ and $(W_3,D)$ .
Under the zeroth order $p$-adic expansion, the number of rational points $\nu_0(W_4),\nu_0(W_3)$ of $W_4$ and $W_3$ are equal, and the complex structure of divisor $D$ has no contribution to $\nu_0(W_4)$.
This has been checked by numerical method for a D-brane system $(W_3,D)$ on quntic and its dual $W_4$ .

This article is organized as follows. 
In Section 2, we  briefly introduce the basic about $p$-adic numbers and calculations of the number of rational points over finite fields. 
In Section 3, we introduce the results in \cite{4} at first.
Then by toric geometry, we derive a general expression for the relation between the number of rational points and periods of two-parameters Calabi-Yau $n$-folds.
In section 4, some corollaries of our results are described by analyzing the structure of the enhanced GKZ-system, and numerical results are given for specific models.
In Section 5, we summarize the main results of this article and some possible future studies.

\section{Finite fields and $p$-adic numbers}
\paragraph{Finite fields}

In order to study the arithmetic properties of Calabi-Yau manifolds, we need to define them over finite fields rather than complex number fields as usual. 
Discussion of finite fields can be found in standard textbook, sufficient introductions also are provided in \cite{4}, but for completeness, this article contains the most essential knowledge.

The structure of a finite field $\mathbb{F}_{p^n}$ for prime number $p$ and $n\in \mathbb Z^+$  is completely determined by the number of its elements $p^n$ . 
The simplest finite field is the integer $\mathrm{mod}\  p$
$$
\mathbb{F}_p=\mathbb{Z}\ \mathrm{mod}\ p=\{0,1,...,p-1\}
$$
In this article, we just discuss finite fields containing prime elements.

In elementary number theory, Fermat's little theorem is well known:
\begin{equation} 
a^{p}=a\ \mathrm{mod}\  p
\end{equation}
In $\mathbb{F}_p$, the calculation of $ \mathrm{mod}\  p$ is automatically included, so this theorem can be expressed as:

\begin{equation} \label{eq:4}
a^{p-1}=\left\{\begin{array}{l}
1,  \ \mathrm{if} \ \  a \neq 0 \\
0, \ \mathrm{if}  \ \ a=0
\end{array}\right.
\end{equation}
This conclusion is important for us to calculate the number of rational points. 

\paragraph{$p$-adic numbers}
In the study of the relation between the number of rational points and periods, one can find that the relation between the complete numbers of rational points and periods is relatively complicated, but if the number of rational points reduced by $\mathrm{mod}\ p^n$ is considered:
\begin{equation} \label{eq:5}
\nu_{n-1}=\nu\  \mathrm{mod}\  p^n 
\end{equation}
the relation become clear. 
Obviously, after $ \mathrm{mod}\  p$ reduction, most of the information about rational points is lost, and $\nu_0$ can only be regarded as a ``zeroth order approximation" of the number of rational points.
In order to recover the complete rational points, we want the following expansion to exist:
\begin{equation} \label{eq:6}
\nu=\nu_0+\nu_1p+\nu_2p^2+...
\end{equation}
Calculating the expansion of order $n$ is equivalent to calculating rational points under $\mathrm {mod}\ p^{n+1}$. 
However, prime $p$ is not a infinitesimal in the general sense. 
In order to abtain the expansion, we introduce the concept of $p$-adic number and redefine the norm. 
Notice that any rational number $r$ can always be written as:
$$
r=\frac{m}{n}=\frac{m_0}{n_0}p^{\alpha}
$$
where $p$ is prime. 
We define the $p$-adic norm of the rational number $r$ as:
$$
\|r\|_p=p^{-\alpha}\ ,\|0\|_p=0
$$
It is easy to verify that it satisfies the definition of norm. 
The $p$-adic norm can be used to complete the rational number field:
$
\mathbb{Q}\to\mathbb{Q}_p
$. 
$\mathbb{Q}_p$ is called a $p$-adic number field, its elements are called $p$-adic number and can be expressed as series:
\begin{equation} \label{eq:10}
A=\sum_{n=n_0}^{\infty}a_np^n\ ;\ a_n\in\mathbb{Z}
\end{equation}
This is exactly the power series we need, the rational points expansioned by $p$. 
After defining the $p$-adic norm, the concept of $p$-adic analysis naturally follows, but we will not discuss it in detail. 
Now, the prime $p$ can be treated as infinitesimal, which allows us to write $\mathrm{mod}\ p$ reduced as:
\begin{equation} \label{eq:11}
\nu_0=\nu\  \mathrm{mod}\  p\Leftrightarrow\nu=\nu_0+o(p)
\end{equation}

\paragraph{The number of rational points}
For algebraic surfaces over $\mathbb F_p$ defined by zeros of polynomials $P(x;\psi)$ with moduli $\psi\in\mathbb F_p$  :
$$
M(\psi)=\{x\in\mathbb{F}_p^5|P(x;\psi)=0\}
$$
the number of rational points are defined as:
\begin{equation} \label{eq:13}
\nu(\psi)=\#\{x\in\mathbb{F}_p^5|P(x;\psi)=0\}
\end{equation}
Of course, the numerical value of rational points can be easily obtained by computer program, but in order to study the relation between the number of rational points and periods, it is necessary to put the problem on $p$-adic number field and calculate the approximation of rational points in each order.
The specific method is to use Fermat's little theorem, $\forall x\in\mathbb{F}_p^5$, we have:
\begin{eqnarray} \label{eq:1}
P(x;\psi)^{p-1}=\left\{\begin{array}{l}
1+o(p),\ \mathrm { if }\ x \notin X_\psi \\
0+o(p),\ \mathrm { if }\ x\in X_\psi
\end{array}\right.
\end{eqnarray}
The count of rational points can then be constructed:
\begin{equation} \label{eq:2}
\nu(\psi)=\sum_{x\in\mathbb{F}_p^5}(1-P(x;\psi)^{p-1})+o(p)
\end{equation}
This expression is equivalent to the zeroth order approximation in the sense of $p$-adic analysis. 
To obtain an exact solution for the number of rational points, we can consider higher-order corrections:
\begin{eqnarray}\nonumber
P(x;\psi)^{p-1}&=&1+o(p) \\\nonumber
P(x;\psi)^{p(p-1)}&=&1+o(p^2)\\\nonumber
...
\end{eqnarray}
This is the main method to be used later in this article.

\section{The number of rational points for Calabi-Yau}

\subsection{The number of rational points for quntic}

For the elliptic curve $X(\tau)$ defined over finite field $\mathbb F_p$ with complex moduli $\tau$, the number of rational points of the curve $X(\tau)$ can be expressed as \cite{2} :
$$
\nu(\tau)=(-1)(-1)^{(p-1)/2}\sum_{n=0}^{\infty}\binom{-1/2}{r}^2\tau^n
$$
This is the same as the expression of the period $\pi (\tau)$ of $X(\tau)$. 
For the quintic Calabi-Yau manifold $M_3(\psi)$ in \cite{4}, the defining equation is:
$$
P(x;\psi)=\sum_{i=1}^{5}x_i^5-5\psi\prod_{i=1}^5 x_i
$$
Where $x\in\mathbb F^5_p$ , and $\psi\in\mathbb F_p$ is the complex moduli.
According to (\ref{eq:2}) in Section 2, one can write the number of rational points of $M_3(\psi)$ as :
\begin{eqnarray}\nonumber
\nu(\psi)
=p^5-\sum_{x\in\mathbb{F}_p^5}\left(\sum_{i=1}^{5}x_i^5-5\psi\prod_{i=1}^5 x_i\right)^{p-1}+o(p)
\end{eqnarray}
the number of rational points can be obtained by simplification:
\begin{eqnarray}
\nu(\psi)=\sum_{n=0}^{[\frac{p}{5}]}
\frac{(5n)!}{(n!)^5}
z^{n}
+o(p)
\end{eqnarray}
where $z=(5\psi)^{-5}$.
This is formally corresponds to the fundamental period of quintic:
\begin{equation} \label{eq:eps}
f_0(\psi)=
\sum_{n=0}^{\infty}
\frac{(5n)!}{(n!)^5}
z^{n} 
\end{equation}
the truncation of the summationthe is the only difference. 
It can be proved by the formula $(ap+b)!=a!b!(p!)^a+o(p^2)$ in $p$-adic analysis that the truncation is generated by transferring the series $\sum_{n=0}^{\infty} a_n z^n $ to the finite field $\mathbb F_p$. Then one can find the equation between the number of rational points and fundamental period of quntic over finite field under zeroth order approximation:
\begin{equation} \label{eq:eps}
\nu(\psi)=\ ^{[p/5]}f_0(\psi)+o(p)
\end{equation}
where $[p/5]$ indicates the above truncation.

There is a similar discussion about first-order corrections of the number of points, and the exact expression of the number of rational points can be guess as:
\begin{equation}\label{eq:1.1}
\nu(\psi)=\sum_{k=0}^{4}
\frac{1}{k!}\left(\frac{p}{1-p}\right)^k
\ ^{(p-1)}f^{(k)}_k(z^{p^4})
\ \mathrm{mod}\  p^5
\end{equation}
Where $f^{(k)}_k$ represents the $k$ logarithmic derivative $(z\frac{d}{dz})^kf_k$ of the $k$-th period $f_k$, and this formula is verified by numerical calculation.

This result relates the important geometric quantity, periods, for the calculation of effective superpotential in physical superstring theory to the number of rational points in arithmetic geometry for the first time, and gives a fairly accurate equation. 
It is still one of the few studies of Calabi-Yau manifolds, the inner spaces of superstrings, over finite fields.

\subsection{The number of rational points for $n$ dimensional two-parameter hypersurface}

In this article, Calabi-Yau manifolds will be described by the language of toric geometry \cite{10}.
Compact Calabi-Yau can be defined as hypersurfaces in toric variety \cite{11}. Consider a Laurent polynomials:
\begin{equation}
f(x)=\sum c_mx^m 
\end{equation}
its monomial corresponds to a convex polyhedron $\Delta$ in lattice space $M_{\mathbb Q}$ :
$$
\Delta(f)\subset M_{\mathbb{Q}}
$$
which is the convex hull of monomial whose coefficients are not equal to zero. 
In fact, this polyhedron $\Delta$ is the same one that defines the toric variety $P_\Delta=\mathrm{Proj}\ S_\Delta$.
The zeros set of each Laurent polynomial corresponds to an affine hypersurface $Z_{f,\Delta}$ in $P_\Delta$ :
$$
Z_{f,\Delta}=\{x\in P_\Delta\ |\ f(x)=0\}
$$
a compact Calabi-Yau manifold can then be defined.
If the polyhedron $\Delta$ is a simplex, or if its vertices give only one linear relation, then the corresponding Calabi-Yau manifold has one-parameter. Otherwise, a multi-parameter object can be obtained.

In \cite{6}, some examples of the number of rational points of Calabi-Yau manifolds are examined in detail by Gauss sum. 
For our purposes, the relation between rational points and periods of multi-parameter hypersurfaces is expected to be given. 
Taking the two-parameter case as an example, for the $n$-dimensional toric variety embedded in the $N+n$-dimensional ambient space, there is a general two-parameter hypersurface equation:
\begin{eqnarray} \label{eq:17}
P(x;\phi,\psi)=\sum_{\bold m\in\Delta}\alpha_{\bold m}x^{\bold m}=\sum_{\bold m\in \Delta^{\prime}}\alpha_{\bold m}x^{\bold m}-\phi Q-\psi U
\end{eqnarray} 
We use Fermat's little theorem (\ref{eq:1}) in Section 2 , instead of the Gauss sum, to compute the number of rational points in zeroth order:
\begin{eqnarray}\nonumber
\nu_0(\phi,\psi)
&=&\sum_{x\in\mathbb{F}_p^{N+n}}(1-(P(x;\phi,\psi))^{p-1})\nonumber\\\nonumber
&=&p^{N+n}-\sum_{x\in\mathbb{F}_p^{N+n}}(\sum_{\bold m\in \Delta^{\prime}}\alpha_{\bold m}x^{\bold m}-\phi Q-\psi U)^{p-1}
\end{eqnarray}
bt polynomial theorem , above equation can be expand as:
\begin{eqnarray}\nonumber
\nu_0(\phi,\psi)
&=&
\sum_{x\in\mathbb{F}_p^{*N+n}}
\sum_{n_1,n_2}
\frac{(p-1)!}{(p-1-n_1-n_2)!n_1!n_2!}
(\sum_{\bold m\in \Delta^{\prime}}
\alpha_{\bold m}x^{\bold m})^{n_1+n_2}(-\phi Q)^{-n_1}(-\psi U)^{-n_2}\nonumber
\end{eqnarray}
considering that $p$ is infinitesimal in $p$-adic analysis, we has the following formula:
\begin{eqnarray}\label{eq:3}
\frac{(p-1)!}{(p-1-n_1-n_2)!}&=&(p-1)(p-2)...(p-1-n_1-n_2)\nonumber\\
&=&(-1)^{n_1+n_1}(n_1+n_2)!+o(p)
\end{eqnarray}
in terms of (\ref{eq:3}), the number of rational points can be reduced to:
\begin{eqnarray}
\nu_0(\phi,\psi)
&=&
\sum_{x\in\mathbb{F}_p^{*N+n}}
\sum_{n_1,n_2}
\frac{(-1)^{n_1+n_1}(n_1+n_2)!}{n_1!n_2!}
(\sum_{\bold m\in \Delta^{\prime}}
\alpha_{\bold m}x^{\bold m})^{n_1+n_2}
(-\phi Q)^{-n_1}(-\psi U)^{-n_2}
\nonumber\\
&=&
\sum_{x\in\mathbb{F}_p^{*N+n}}
\sum_{n_1,n_2}
\frac{(n_1+n_2)!}{n_1!n_2!}\sum_{\sum n_{\bold m}=n_1+n_2}\frac{(n_1+n_2)!}{\prod_{\bold m}n_{\bold m}!}
\prod_{\bold m\in\Delta^{\prime}}(\alpha_{\bold m}x^{\bold m})^{n_{\bold m}}\frac{1}{\phi^{n_1}\psi^{n_2}Q^{n_1}U^{n_2}}\nonumber\\
&=&
\sum_{n_1,n_2}
\frac{[(n_1+n_2)!]^2}{n_1!n_2!}
\frac{1}{\phi^{n_1}\psi^{n_2}}
\sum_{\sum n_{\bold m}=n_1+n_2}
\prod_{\bold m\in\Delta^{\prime}}
\frac{\alpha^{n_{\bold m}}_{\bold m}}{n_{\bold m}!}
\sum_{x\in\mathbb{F}_p^{*N+n}}
x^{\sum_{\bold m\in\Delta^{\prime}}{n_{\bold m}}\bold m-n_1\bold q-n_2 \bold u}\nonumber
\end{eqnarray}
further expansion by polynomial theorem, finally we have:
\begin{eqnarray}
\nu_0(\phi,\psi)
&=&
\sum_{x\in\mathbb{F}_p^{*N+n}}
\sum_{n_1,n_2}
\frac{(n_1+n_2)!}{n_1!n_2!}\sum_{\sum n_{\bold m}=n_1+n_2}\frac{(n_1+n_2)!}{\prod_{\bold m}n_{\bold m}!}
\prod_{\bold m\in\Delta^{\prime}}(\alpha_{\bold m}x^{\bold m})^{n_{\bold m}}\frac{1}{\phi^{n_1}\psi^{n_2}Q^{n_1}U^{n_2}}\nonumber\\
&=&
\sum_{n_1,n_2}
\frac{[(n_1+n_2)!]^2}{n_1!n_2!}
\frac{1}{\phi^{n_1}\psi^{n_2}}
\sum_{\sum n_{\bold m}=n_1+n_2}
\prod_{\bold m\in\Delta^{\prime}}
\frac{\alpha^{n_{\bold m}}_{\bold m}}{n_{\bold m}!}
\sum_{x\in\mathbb{F}_p^{*N+n}}
x^{\sum_{\bold m\in\Delta^{\prime}}{n_{\bold m}}\bold m-n_1\bold q-n_2 \bold u}\nonumber\\
\end{eqnarray}
This is the zeroth order expression of the number of rational points of $n-1$ dimensional two-parameter hypersurfaces.

\subsection{The fundamental period for $n$ dimensional two-parameter hypersurface}

For the hypersurface $Z_{f,\Delta}$ in toric variety, such as the mirror manifold $W_3$ in usual, which defined by the dual polyhedron $\Delta^*$ in mirror symmetry, the period integral is defined as:
\begin{equation} \label{eq:4}
\Pi_i= \int_{\gamma_i} \frac{1}{f(x)} \prod_{j=1}^{n} \frac{dx_j}{x_j}
\end{equation}
These periods, as solutions of the Picard-Fuchs equation, can be obtained from the GKZ-generalized hypergeometric system related to $\Delta^*$ \cite{12}, \cite{13}.
But in order to compare the number of rational points in zeroth order $\nu_0(\phi,\psi)$ , we use the definition (\ref{eq:4}) to calculate the fundamental period $\varpi_0(\phi,\psi)$ of two-parameter hypersurface directly :
\begin{eqnarray}\label{5}
\varpi_0(\phi,\psi)
&=&
\frac{C}{(2\pi i)^{N+n}}\int\frac{d^{N+n}x}{-P(\phi,\psi)}
\end{eqnarray}
Plug the polynomial equation (\ref{eq:17}) into (\ref{5}) :
\begin{eqnarray}
\varpi_0(\phi,\psi)
&=&
\frac{C}{(2\pi i)^{N+n}}\int\frac{d^{N+n}x}
{\phi Q+\psi U-\sum_{\bold m\in \Delta^{\prime}}\alpha_{\bold m}x^{\bold m}}\nonumber\\
&=&
\frac{C}{(2\pi i)^{N+n}}\int
\frac{d^{N+n}x}{\phi Q+\psi U}
\left(1-
\frac{\sum_{\bold m\in \Delta^{\prime}}\alpha_{\bold m}x^{\bold m}}
{\phi Q+\psi U}\right)^{-1}\nonumber
\end{eqnarray}
consider the power series expansion of the integrand:
\begin{eqnarray}
&&\left(1-\frac{\sum_{\bold m\in \Delta^{\prime}}\alpha_{\bold m}x^{\bold m}}{\phi Q+\psi U}\right)^{-1} \nonumber\\
&&=\sum_{n_1+n_2}
\frac{1}{(n_1+n_2)!}
\sum_{n_1,n_2}\nonumber
\frac{(n_1+n_2)!}{n_1!n_2!}
\frac{1}{{(Q\phi)}^{n_1}{(U\psi)}^{n_2}}
\partial^{n_1}_{1/{Q\phi}}
\partial^{n_2}_{1/{U\psi}}
\left(1-\frac{\sum_{\bold m\in \Delta^{\prime}}\alpha_{\bold m}x^{\bold m}}
{\phi Q+\psi U}\right)^{-1}\nonumber\\
&&=
\sum_{n_1,n_2}
\frac{1}{n_1!n_2!}
\frac{1}{\phi^{n_1}\psi^{n_2}}
(n_1+n_2)! \nonumber
(\sum_{\bold m\in \Delta^{\prime}}\alpha_{\bold m}x^{\bold m})^{n_1+n_2}
Q^{-n_1}
P^{-n_2} 
\end{eqnarray}
expand by polynomial theorem:
\begin{eqnarray}
&&
\sum_{n_1,n_2}
\frac{1}{n_1!n_2!}
\frac{1}{\phi^{n_1}\psi^{n_2}}
(n_1+n_2)!
\sum_{\sum n_{\bold m}=n_1+n_2}
\frac{(n_1+n_2)!}{\prod_{\bold m}n_{\bold m}}
\prod_{\bold m\in\Delta^{\prime}}
\frac{(\alpha_{\bold m}x^{\bold m})^{n_{\bold m}}}{n_{\bold m}}
Q^{-n_1}
P^{-n_2}
\nonumber\\
&&=
\sum_{n_1,n_2}
\frac{[(n_1+n_2)!]^2}{n_1!n_2!}
\frac{1}{\phi^{n_1}\psi^{n_2}}
\sum_{\sum n_{\bold m}=n_1+n_2}
\prod_{\bold m\in\Delta^{\prime}}
\frac{(\alpha_{\bold m}x^{\bold m})^{n_{\bold m}}}{n_{\bold m}}
Q^{-n_1}
P^{-n_2}\nonumber
\end{eqnarray}
Finally, we find that the explicit expression of the fundamental period and the number of rational points in zeroth order approximation, namely:
\begin{eqnarray}
\nu_0(\phi,\psi)
&=&
\sum_{n_1,n_2}
\frac{[(n_1+n_2)!]^2}{n_1!n_2!}
\frac{1}{\phi^{n_1}\psi^{n_2}}
\sum_{\sum n_{\bold m}=n_1+n_2}
\prod_{\bold m\in\Delta^{\prime}}
\frac{\alpha^{n_{\bold m}}_{\bold m}}{n_{\bold m}!}
\\
&&\times
\sum_{x\in\mathbb{F}_p^{*N+n}}
x^{\sum_{\bold m\in\Delta^{\prime}}{n_{\bold m}}\bold m-n_1\bold q-n_2 \bold u}
\nonumber\\
\varpi_0(\phi,\psi)
&=&
\sum_{n_1,n_2}
\frac{[(n_1+n_2)!]^2}{n_1!n_2!}
\frac{1}{\phi^{n_1}\psi^{n_2}}
\sum_{\sum n_{\bold m}=n_1+n_2}
\prod_{\bold m\in\Delta^{\prime}}
\frac{\alpha_{\bold m}^{n_{\bold m}}}{n_{\bold m}!}
\\
&&\times
\frac{C}{(2\pi i)^{N+n}}
\int
\frac{d^{N+n}x}{\phi Q+\psi U}
x^{\sum_{\bold m\in\Delta^{\prime}}{n_{\bold m}}\bold m-n_1\bold q-n_2 \bold u}
\nonumber
\end{eqnarray}
Since the same restricts are given by $x$-sum and integral in the formula above \cite{4} :
\begin{eqnarray}
\sum_{\bold m\in\Delta^{\prime}}{n_{\bold m}}\bold m=n_1\bold q+n_2 \bold u
\end{eqnarray}
It means that we prove that the equation of rational points and fundamental period still holds for $n-1$ dimensional two-parameter hypersurface under zeroth order approximation:
\begin{eqnarray} \label{eq:20}
\nu(\phi,\psi)=\ ^{(p-1)}\varpi_0(\phi,\psi)+o(p)
\end{eqnarray}

\section{The number of rational points for 4-fold in F-theory}

\subsection{The type II/F-theory duality}

In toric geometry, one way to construct a multi-parameter Calabi-Yau manifold is to embed the polyhedron $\Delta$ in a 4D lattic space corresponding to the one-parameter Calabi-Yau 3-fold $W_3$ into the 5D lattic space. 
Then add extra vertices $\{\rho_i\}$ to make it an enhanced polyhedron $\tilde \Delta$ \cite{14}.
This means that additional linear relations among vertices are introduced and therefore additional complex moduli parameters are provided.

The above configuration is called the type II/F-theory duality \cite{9} , $\tilde \Delta$ relates two completely different geometry. 
One is the Calabi-Yau 4-fold $W_4$ which is related to F-theory compactification.
$W_4$ is the multi-parameter manifold obtained by elliptic fibration of $W_3$  \cite{15}.
The other is the manifold pair $(W_3,E)$, open strings is compact on $W_3$ , and end on the D-brane wrapping the submanifold $E$.
The periods satisfies the same GKZ-system given by the same enhanced polyhedron $\tilde\Delta$, therefore, although the relative periods $\hat\Pi(W_3,D)$ and the absolute periods $\Pi(W_4)$ has different geometry meaning, the same low-energy effective theory can be given. 
Here $D$ is a divisor on $W_3$ and parameterizes the deformation of D-brane $E$ . 

For the charge vector $l_i(\Delta)$ given by the linear relation $\sum_\Delta l_i(\Delta)v_i=0$ of the vertex $v_i\in \Delta$, since the linear relation is maintained in the embedding $v_i\mapsto u_i=(v_i,0)$ from 4D to 5D , the same charge vector $\hat l_i(\tilde\Delta)=(l_i(\Delta),0^k)$ holds for extended polyhedra $\tilde \Delta$ , where $k=\#\{\rho_i\}$ .
In GKZ-system, it means the same Picard-Fuchs operators:
\begin{eqnarray}
\{\prod_{l_j^{a}>0}(\prod_{i=0}^{l_j^{a}-1}(\theta_{j}-i))-
\prod_{i=1}^{|l_{0}^{a}|}(i-
|l_{0}^{a}|-\theta_{0}) \prod_{l_j^{a}<0,j \neq 0}(\prod_{i=0}^{|l_{j}^{a} \mid-1}(\theta_{j}+|l_{j}^{a}|-i)) z_{a}\}\Pi(z)=0
\end{eqnarray}
where $\theta_j=z_j\frac{d}{d z_j}$ .
Note that all periods of $W_3$ consists of a subset of the periods of $W_4$. 
In particular, the two manifolds have the same fundamental period, $\varpi_0 (W_3)=\varpi_0 (W_4)$.
By the conclusion of Section 3 , the relation between complex structures of two manifolds will be expressed on the number of rational points over finite fields.

\subsection{The corollary}
By the formula (\ref{eq:20}) at the end of Section 3, we can immediately obtain some corollaries about the relation between rational points of $W_3(\psi)$ and $W_4(\phi,\psi)$ over finite fields $\mathbb F_p$. The one-parameter polynomials and two-parameter polynomials 
$$
P_3(x;\psi)=\sum_{\bold m\in\Delta^*}\alpha_{\bold m}x^{\bold m}
$$
$$
P_4(x;\phi,\psi)=\sum_{\bold m\in\tilde\Delta^*}\alpha_{\bold m}x^{\bold m}
$$
given by the polyhedron $\Delta^*$ and its enhanced $\tilde \Delta^*$ define 3D and 4D Calabi-Yau hypersurfaces are dual space in the sense of type II/F-theory duality and have the same fundamental period $\varpi_0=\ ^{[p/n]}f_0$ .
Therefore, their the number of rational points are:
\begin{eqnarray} \label{eq:13}
\nu(P_3)&=&\#\{x\in\mathbb{F}_p^n|P_3(x;\psi)=0\ \mathrm{mod}\ p\}\nonumber\\
\nu(P_4)&=&\#\{x\in\mathbb{F}_p^{\tilde n}|P_4(x;\phi,\psi)=0\ \mathrm{mod}\ p\}\nonumber
\end{eqnarray}
Where $n$ is the number of vertices of $\Delta$ , $\tilde n$ is the number of vertices of $\tilde \Delta$ dual of $\tilde \Delta^*$.
Since they are equal to the same fundamental period in the zeroth order approximation, we can write $\nu_0(P_3)=\nu_0(P_4)$, or equivalent:
\begin{eqnarray}\label{eq:22}
\nu(P_3)=\nu(P_4)\ \mathrm{mod}\ p
\end{eqnarray}
This result relate the arithmetic properties of two toric varieties with different dimensions, and is not obvious in mathematics.
But a simple proof is given by type II/F-theory duality in physics.
In fact, the whole reasoning does not depends on the dimension of toric hypersurfaces, so (\ref{eq:22}) can be generalized to $n$-dimension.

One can also observe the effect of elliptic fibration process from 3- to 4-dimensional manifold on the number of rational points over finite field.
In toric geometry, elliptic fibration is determined by the extra vertices $\{\rho_i\}$, and thus is  related to the divisor $D$ on the Calabi-Yau 3-fold $W_3$.  
However, over finite fields, the different geometry of $D$ , or the different elliptic fibration, has no influence on the zeroth order $p$-adic expansion of the number of rational points. 

\subsection{Example: Quintic and the dual 4-fold}

For specific models under the type II/F-theory duality, equation (\ref{eq:22}) can be tested numerically.
For example, one can consider the vertices of the polyhedron $\Delta$ and its dual $\Delta^*$ are:
\begin{table}[h!]
\centering
\begin{tabular}{|c|c|c|c|}
\hline
\multirow{5}{1cm}{$\Delta$}&$ v_1=(-1, -1, -1, -1)$&\multirow{5}{1cm}{$\Delta^*$}&$ v_1^*=(-1, -1, -1, -1)$\\
\cline{2-2}\cline{4-4}
& $v_2= (-1, -1, -1, 4)$& & $v_2^*= (0, 0, 0, 1)$\\
\cline{2-2}\cline{4-4}
& $v_3=(-1, -1, 4, -1)$& & $v_3^*=(0, 0, 1, 0)$\\
\cline{2-2}\cline{4-4}
& $v_4=(-1,4, -1, -1)$& & $v_4^*=(0, 1, 0, 0)$\\
\cline{2-2}\cline{4-4}
& $v_5=(4, -1, -1, -1)$& & $v_5^*=(1, 0, 0, 0)$\\
\hline
\end{tabular}
\caption{the vertices of $\Delta$ and $\Delta^*$}
\end{table}

Calabi-Yau hypersurface corresponds to $\Delta^*$ is given by the formula:
\begin{eqnarray}\label{eq:23}
P(\Delta^*)=\sum_{v_j\in\Delta^*}a_i\prod_{v_i\in\Delta}x_j^{\langle v,v_i^*\rangle+1}
\end{eqnarray}
In this case it is known as the quintic :
$$
P_3(\Delta^*)=x_1^5 + x_2^5 + x_3^5 + x_4^5 + x_5^5+ a_0 x_1 x_2 x_3 x_4 x_5 
$$

Embedding $\Delta^*$ into the 5D lattic space and specify the extra vertices corresponding to the divisor $D$ as:
$$
\rho_1=(0, 0, 0, 0,1)\ ,\ 
\rho_2=(-1, -1, -1, -1,1)
$$
\newpage
Then there are extended polyhedra $\tilde\Delta^*$ and their duals $\tilde\Delta$ :
\ 
\begin{table}[h!]
\centering
\begin{tabular}{|c|c|c|c|}
\hline
\multirow{9}{1cm}{$\tilde\Delta$}&$ v_1=(-1, -1, -1, -1, -1)$&\multirow{9}{1cm}{$\tilde\Delta^*$}&$ v_1^*=(-1, -1, -1, -1,0)$\\
\cline{2-2}\cline{4-4}
& $v_2= (-1, -1, -1, 3, -1)$& & $v_2^*= (0, 0, 0, 1,0)$\\
\cline{2-2}\cline{4-4}
& $v_3=(-1, -1, -1, 4, 0)$& & $v_3^*=(0, 0, 1, 0,0)$\\
\cline{2-2}\cline{4-4}
& $v_4=(-1, -1, 3, -1, -1)$& & $v_4^*=(0, 1, 0, 0,0)$\\
\cline{2-2}\cline{4-4}
& $v_5=(-1, -1, 4, -1, 0)$& & $v_5^*=(1, 0, 0, 0,0)$\\
\cline{2-2}\cline{4-4}
& $v_6=(-1, 3, -1, -1, -1)$& & $\rho_1=(0, 0, 0, 0,1)$\\
\cline{2-2}\cline{4-4}
& $v_7=(-1, 4, -1, -1, 0)$& & $\rho_2=(-1, -1, -1, -1,1)$\\
\cline{2-2}\cline{4-4}
& $v_8=(3, -1, -1, -1, -1)$& & \\
\cline{2-2}\cline{4-4}
& $v_9=(4, -1, -1, -1, 0)$& & \\
\hline
\end{tabular}
\caption{$\tilde\Delta$ and $\tilde\Delta^*$ 's vertices}
\end{table}

The defining equation of to the Calabi-Yau 4-folds of F-theory are also given by (\ref{eq:23}):
$$
P_4(\tilde\Delta^*)=x_1^4 + x_2^4 x_3^5 + x_4^4 x_5^5 + x_6^4 x_7^5+ x_8^4 x_9^5 + 
 x_1^5 x_2 x_4 x_6 x_8 +b_0 x_3 x_5 x_7 x_9 + 
 a_0 x_1 x_2 x_3 x_4 x_5 x_6 x_7 x_8 x_9 
$$
By selecting different complex moduli parameters $a_0,b_0$, we calculate the number of rational points for the first five prime numbers, see Appendix A. 
For all $a_0,b_0$ and prime $p$, the calculation results show that equation (\ref{eq:22}) is valid.

\section{Summary and outlook}
The main motivation of this article is to generalize the results of \cite{4} to more general cases. 
In Section 3, we use toric geometry to give a relatively general proof for the proposition that over $p$-adic field, the fundamental period of two-parameter Calabi-Yau $n$-folds is equal to the number of rational points in the zeroth order approximation. 

The 4D Calabi-Yau manifolds corresponding to F-theory is an example of this case, so our result can be used to study open string and type II/F-theory duality over finite fields. 
In particular, by the formula (\ref{eq:20}) , we deduce that the 3D and 4D Calabi-Yau manifolds of type II/F-theory duality have the same number of rational points in zeroth order over a finite field.
Taking quintic hypersurface as an example, we verify this conclusion under different complex moduli parameters.

For manifold pair $(W_3, E)$ or $(W_3,D)$ that is interesting for open string compactification, we
expect that similar conclusion holds for Calabi-Yau manifold
$M$ and divisor $D$ respectively.
To verify this, a method based on type II/F-theory duality will be presented in the follow-up work.
This is not only the application of results in this article, but also the evidence of type II/F-theory duality over finite fields.

As the formula (\ref{eq:20}) in Section 3 is only accurate to zeroth order, its improvement is also what we need to do in the next step. 
In the case of Calabi-Yau manifolds with more than one moduli, the periods given by different charge vectors form several subsets.
For the higher-order correction of rational points, how to combine the periods belonging to different subsets is the primary problem we need to solve.
Processing of higher-order corrections will help us to see the effect of extra periods, or "opening periods" in open string case, on the number of rational points.
Clarification of this structure is necessary to analyze the properties of rational points in elliptic fibration and type II/F-theory duality. 

In addition, it might make sense to consider the mirror symmetry of $\zeta$ function in the open string case.

\newpage
\begin{appendices}
\section{numerical results}

\begin{table}[h!]
\centering
\begin{tabular}{|c|c|c|c|c|c|c|}
\hline
 & $a_0=0$ & $b_0=0$ & $b_0=1$ & $b_0=2$ & $b_0=3$ & $b_0=4$ \\
\hline
\multirow{3}{1.2cm}{$p=2$} & $\nu(P_3)$ & 16 &16 &&& \\

 & $\nu(P_4)$ & 256 &254 &&& \\

 & $\nu_0(P_3),\nu_0(P_4)$ & (0,0) & (0,0)&&&\\ 
\cline{1-7}

\multirow{3}{1.2cm}{$p=3$} & $\nu(P_3)$ & 81 & 81&81 && \\

 & $\nu(P_4)$ & 6561 &6609 &6465 && \\

 & $\nu_0(P_3),\nu_0(P_4)$ & (0,0) &(0,0) &(0,0) &&\\ 
\cline{1-7}

\multirow{3}{1.2cm}{$p=5$} & $\nu(P_3)$ & 625 & 625&625 &625 &625 \\

 & $\nu(P_4)$ & 390625 &386785 &389345 &394465 &390625 \\

 & $\nu_0(P_3),\nu_0(P_4)$ & $(0,0)$&(0,0) &(0,0) &(0,0) &(0,0)\\ 
\cline{1-7}

\multirow{3}{1.2cm}{$p=7$} & $\nu(P_3)$ & 2401 &2401 &2401 &2401 &2401 \\

 & $\nu(P_4)$ & 5764801 &5810161 &5846449 &5755729 &5665009 \\

 & $\nu_0(P_3),\nu_0(P_4)$ & $(0,0)$ &(0,0) &(0,0) & (0,0)&(0,0)\\ 
\cline{1-7}

\multirow{3}{1.2cm}{$p=11$} & $\nu(P_3)$ & 19251 &19251 & 19251& 19251&19251 \\

 & $\nu(P_4)$ & 215248881 &214038881 &214478881 &214038881 &214038881 \\

 & $\nu_0(P_3),\nu_0(P_4)$ & $(1,1)$ &(1,1) &(1,1) &(1,1) &(1,1)\\ 
\cline{1-7}

\hline
\end{tabular}
\caption{The number of rational points of $P_3$ and $P_4$ when $a_0=0$}
\end{table}
\begin{table}[h!]
\centering
\begin{tabular}{|c|c|c|c|c|c|c|}
\hline
 & $a_0=1$ & $b_0=0$ & $b_0=1$ & $b_0=2$ & $b_0=3$ & $b_0=4$ \\
\hline
\multirow{3}{1.2cm}{$p=2$} & $\nu(P_3)$ & 17 & 17 &&& \\

 & $\nu(P_4)$ &255 & 255 &&&\\

 & $\nu_0(P_3),\nu_0(P_4)$ &(1,1) & $(1,1)$ &&&\\ 
\cline{1-7}

\multirow{3}{1.2cm}{$p=3$} & $\nu(P_3)$ &73 & 73 & 73&& \\

 & $\nu(P_4)$ &6625 & 6481 &6529 && \\

 & $\nu_0(P_3),\nu_0(P_4)$ &(1,1) & (1,1) & (1,1)&&\\ 
\cline{1-7}

\multirow{3}{1.2cm}{$p=5$} & $\nu(P_3)$ &681 & 681 &681 &681 &681 \\

 & $\nu(P_4)$ &387041 & 401121 &385761 &390881 &387041 \\ 

 & $\nu_0(P_3),\nu_0(P_4)$ &(1,1) & $(1,1)$ &(1,1) & (1,1)&(1,1)\\ 
\cline{1-7}

\multirow{3}{1.2cm}{$p=7$} & $\nu(P_3)$ &2191 & 2191 &2191 &2191 &2191 \\

 & $\nu(P_4)$ &5810161 & 5746657 &5710369 &5782945 &5710369 \\

 & $\nu_0(P_3),\nu_0(P_4)$ &(0,0) & (0,0) &(0,0) &(0,0) &(0,0)\\ 
\cline{1-7}

\multirow{3}{1.2cm}{$p=11$} & $\nu(P_3)$ &25501 & 25501 &25501 &25501 &25501 \\

 & $\nu(P_4)$ &214498881 & 213288881 &214828881 &213728881 &215488881 \\

 & $\nu_0(P_3),\nu_0(P_4)$ &(3,3) & $(3,3)$ &(3,3) &(3,3) &(3,3)\\ 
\cline{1-7}

\hline
\end{tabular}
\caption{The number of rational points of $P_3$ and $P_4$ when $a_0=1$}
\end{table}
\begin{table}[h!]
\centering
\begin{tabular}{|c|c|c|c|c|c|c|}
\hline
 & $a_0=2$ & $b_0=0$ & $b_0=1$ & $b_0=2$ & $b_0=3$ & $b_0=4$ \\
\hline

\multirow{3}{1.2cm}{$p=3$} & $\nu(P_3)$ &91 &91 &91 &&\\

 & $\nu(P_4)$ &6481 &6529 &6625 &&\\

 & $\nu_0(P_3),\nu_0(P_4)$ &(1,1) &(1,1) &(1,1)&&\\ 
\cline{1-7}

\multirow{3}{1.2cm}{$p=5$} & $\nu(P_3)$ &641 &641 &641 &641 &641 \\

 & $\nu(P_4)$ &389601 &385761 &393441 &393441 &389601 \\

 & $\nu_0(P_3),\nu_0(P_4)$ &(1,1) &(1,1) &(1,1) &(1,1) &(1,1)\\ 
\cline{1-7}

\multirow{3}{1.2cm}{$p=7$} & $\nu(P_3)$ &2407 &2407 &2407 &2407 &2407 \\
 & $\nu(P_4)$ &5763505 &5808865 &5700001 &5763505 &5808865 \\

 & $\nu_0(P_3),\nu_0(P_4)$ &(6,6) &(6,6) &(6,6) &(6,6) &(6,6)\\ 
\cline{1-7}

\multirow{3}{1.2cm}{$p=11$} & $\nu(P_3)$ &33001 &33001 &33001 &33001 &33001 \\

 & $\nu(P_4)$ &214038881 &212828881 &215248881 &215028881 &214588881 \\

 & $\nu_0(P_3),\nu_0(P_4)$ &(1,1) &(1,1) &(1,1) &(1,1) &(1,1)\\ 
\cline{1-7}

\hline
\end{tabular}
\caption{The number of rational points of $P_3$ and $P_4$ when $a_0=2$}
\end{table}
\begin{table}[h!]
\centering
\begin{tabular}{|c|c|c|c|c|c|c|}
\hline
 & $a_0=3$ & $b_0=0$ & $b_0=1$ & $b_0=2$ & $b_0=3$ & $b_0=4$ \\
\hline

\multirow{3}{1.2cm}{$p=5$} & $\nu(P_3)$ &561 &561 &561 &561 &561 \\

 & $\nu(P_4)$ &394721 &390881 &393441 &378081 &394721 \\

 & $\nu_0(P_3),\nu_0(P_4)$ &(1,1) &(1,1) &(1,1) &(1,1) &(1,1)\\ 
\cline{1-7}

\multirow{3}{1.2cm}{$p=7$} & $\nu(P_3)$ &2371 &2371 &2371 &2371 &2371 \\

 & $\nu(P_4)$ &5771281 &5671489 &5671489 &5834785 &5852929 \\

 & $\nu_0(P_3),\nu_0(P_4)$ &(5,5) &(5,5) &(5,5) &(5,5) &(5,5)\\ 
\cline{1-7}

\multirow{3}{1.2cm}{$p=11$} & $\nu(P_3)$ &25501 &25501 &25501 &25501 &25501 \\

 & $\nu(P_4)$ &214498881 &215488881 &214828881 &213288881 &213728881 \\

 & $\nu_0(P_3),\nu_0(P_4)$ &(3,3) &(3,3) &(3,3) &(3,3) &(3,3)\\ 
\cline{1-7}

\hline
\end{tabular}
\caption{The number of rational points of $P_3$ and $P_4$ when $a_0=3$}
\end{table}
\begin{table}[h!]
\centering
\begin{tabular}{|c|c|c|c|c|c|c|}
\hline
 & $a_0=4$ & $b_0=0$ & $b_0=1$ & $b_0=2$ & $b_0=3$ & $b_0=4$ \\
\hline

\multirow{3}{1.2cm}{$p=5$} & $\nu(P_3)$ &621 &621 &621 &621 &621 \\

 & $\nu(P_4)$ &390881 &387041 &389601 &394721 &389601 \\

 & $\nu_0(P_3),\nu_0(P_4)$ &(1,1) &(1,1) &(1,1) &(1,1) &(1,1)\\ 
\cline{1-7}

\multirow{3}{1.2cm}{$p=7$} & $\nu(P_3)$ &2461 &2461 &2461 &2461 &2461 \\

 & $\nu(P_4)$ &5751841 &5833489 &5833489 &5751841 &5688337 \\

 & $\nu_0(P_3),\nu_0(P_4)$ &(4,4) &(4,4) &(4,4) &(4,4) &(4,4)\\ 
\cline{1-7}

\multirow{3}{1.2cm}{$p=11$} & $\nu(P_3)$ &25501 &25501 &25501 &25501 &25501 \\

 & $\nu(P_4)$ &214498881 &213728881 &213728881 &215488881 &213288881 \\

 & $\nu_0(P_3),\nu_0(P_4)$ &(3,3) &(3,3) &(3,3) &(3,3) &(3,3)\\ 
\cline{1-7}

\hline
\end{tabular}
\caption{The number of rational points of $P_3$ and $P_4$ when $a_0=4$}
\end{table}
\end{appendices}
\newpage
\ 
\newpage
\

\end{document}